\documentclass[11pt]{article}
\usepackage{times}
\usepackage{sidecap}
\usepackage{geometry}
\usepackage{floatrow}
\usepackage{float}
\usepackage[draft]{hyperref}
\hypersetup{hidelinks}
\usepackage{graphics,graphicx}
\usepackage{subcaption}
\usepackage{epsf,wrapfig}

\usepackage{natbib,bm}

\usepackage{paralist}

\DeclareGraphicsExtensions{.jpg,.pdf,.png,.eps}

\usepackage{caption}
\usepackage{xcolor}
\usepackage{booktabs}
\usepackage{microtype}

\usepackage{tcolorbox}
\tcbuselibrary{breakable}

\bibpunct{(}{)}{;}{a}{}{,}

\usepackage{soul}

\definecolor{darkcyan}{rgb}{0.0, 0.0, 0.7}

\definecolor{boxcolor}{RGB}{193, 229, 255}
\definecolor{figboxcolor}{RGB}{193, 229, 255}
\definecolor{boxcolorgray}{RGB}{227, 227, 227}
\definecolor{darkcyan}{rgb}{0.0, 0.0, 0.7}

\sethlcolor{boxcolorgray}

\def\gtrsim{\lower 2pt \hbox{$\, \buildrel {\scriptstyle >}\over
{\scriptstyle \sim}\,$}}
\def\lesssim{\lower 2pt \hbox{$\, \buildrel {\scriptstyle <}\over
{\scriptstyle \sim}\,$}}

\def\approxlt{\lower.2em\hbox{$\buildrel < \over \sim$}}
\def\approxgt{\lower.2em\hbox{$\buildrel > \over \sim$}}

\def\be{\begin{equation}}
\def\ee{\end{equation}}
\def\bea{\begin{eqnarray*}}
\def\eea{\end{eqnarray*}}

\def\jref@jnl#1{{\rm#1}}
\def\aj{\jref@jnl{AJ}}
\def\araa{\jref@jnl{ARA\&A}}
\def\apj{\jref@jnl{ApJ}}
\def\apjl{\jref@jnl{ApJ}}
\def\apjs{\jref@jnl{ApJS}}
\def\ao{\jref@jnl{Appl.~Opt.}}
\def\apss{\jref@jnl{Ap\&SS}}
\def\aap{\jref@jnl{A\&A}}
\def\aapr{\jref@jnl{A\&A~Rev.}}
\def\aaps{\jref@jnl{A\&AS}}
\def\azh{\jref@jnl{AZh}}
\def\baas{\jref@jnl{BAAS}}
\def\jrasc{\jref@jnl{JRASC}}
\def\memras{\jref@jnl{MmRAS}}
\def\mnras{\jref@jnl{MNRAS}}
\def\pra{\jref@jnl{Phys.~Rev.~A}}
\def\prb{\jref@jnl{Phys.~Rev.~B}}
\def\prc{\jref@jnl{Phys.~Rev.~C}}
\def\prd{\jref@jnl{Phys.~Rev.~D}}
\def\pre{\jref@jnl{Phys.~Rev.~E}}
\def\prl{\jref@jnl{Phys.~Rev.~Lett.}}
\def\pasp{\jref@jnl{PASP}}
\def\pasj{\jref@jnl{PASJ}}
\def\qjras{\jref@jnl{QJRAS}}
\def\skytel{\jref@jnl{S\&T}}
\def\solphys{\jref@jnl{Sol.~Phys.}}
\def\sovast{\jref@jnl{Soviet~Ast.}}
\def\ssr{\jref@jnl{Space~Sci.~Rev.}}
\def\zap{\jref@jnl{ZAp}}
\def\nat{\jref@jnl{Nature}}
\def\iaucirc{\jref@jnl{IAU~Circ.}}
\def\aplett{\jref@jnl{Astrophys.~Lett.}}
\def\apspr{\jref@jnl{Astrophys.~Space~Phys.~Res.}}
\def\bain{\jref@jnl{Bull.~Astron.~Inst.~Netherlands}}
\def\fcp{\jref@jnl{Fund.~Cosmic~Phys.}}
\def\gca{\jref@jnl{Geochim.~Cosmochim.~Acta}}
\def\grl{\jref@jnl{Geophys.~Res.~Lett.}}
\def\jcp{\jref@jnl{J.~Chem.~Phys.}}
\def\jgr{\jref@jnl{J.~Geophys.~Res.}}
\def\jqsrt{\jref@jnl{J.~Quant.~Spec.~Radiat.~Transf.}}
\def\memsai{\jref@jnl{Mem.~Soc.~Astron.~Italiana}}
\def\nphysa{\jref@jnl{Nucl.~Phys.~A}}
\def\physrep{\jref@jnl{Phys.~Rep.}}
\def\physscr{\jref@jnl{Phys.~Scr}}
\def\planss{\jref@jnl{Planet.~Space~Sci.}}
\def\procspie{\jref@jnl{Proc.~SPIE}}

\def\farcs{\hbox{$.\!\!^{\prime\prime}$}}

\def\micron{\hbox{$\mu$m}}

\def\ion#1#2{#1$\;${\small\rm\@Roman{#2}}\relax}

\newcount\lecurrentfam
\def\LaTeX{\lecurrentfam=\the\fam \leavevmode L\raise.42ex
\hbox{$\fam\lecurrentfam\scriptstyle\kern-.3em A$}\kern-.15em\TeX}
\def\sizrpt{
(\fontname\the\font): em=\the\fontdimen6\font, ex=\the\fontdimen5\font
\typeout{
(\fontname\the\font): em=\the\fontdimen6\font, ex=\the\fontdimen5\font
}}
\definecolor{dark}{rgb}{ 0.34, 0.09, 0.49}

\usepackage{fancyhdr,url}
\usepackage{transparent}
\usepackage{tablefootnote}
\usepackage{mathpazo}

\begin{document}

\noindent
\begin{raggedright}
{\Huge\bfseries Has JWST Seen Everything?
\\[0.25em]  \Large
Hunting for ``NIRCam-dark'' Dusty Star-Forming Galaxies at $z \simeq 6 - 9$ with ALMA2040
\par}\end{raggedright}

\vspace{1em}

\noindent
Fengwu Sun$^{1,\ast}$,
Bel\'{e}n Alcalde Pampliega$^2$,
Hiddo Algera$^3$,
Manuel Aravena$^4$,
Tom Bakx$^5$,
Karina I.\ Caputi$^6$,
Pierre Cox$^7$,
Elisabete da Cunha$^8$,
Andreas Faisst$^9$,
Yoshinobu Fudamoto$^{10}$,
Carlotta Gruppioni$^{11}$,
Mahmoud Hamed$^{12}$,
Jacqueline Hodge$^{13}$,
Patrick S.\ Kamieneski$^5$,
Adarsh Kumar$^{14}$,
Ivanna Langan$^{12}$,
Andrea Lapi$^{15}$,
Umberto Maio$^{16}$,
Chiara Mazzucchelli$^4$,
Dieu D.\ Nguyen$^{17}$,
Mamta Pandey-Pommier$^{18}$,
Dominik A.\ Riechers$^{19,20}$,
Irene Shivaei$^{12}$,
Sabine Thater$^{21}$

\vspace{1em}
{\footnotesize
\noindent $^\ast$\textbf{Corresponding author: \href{mailto:sunfengwu@westlake.edu.cn}{sunfengwu@westlake.edu.cn}}\\
}

\begin{tcolorbox}[breakable,boxsep=3pt,left=0pt,right=0pt,top=1pt,bottom=1pt,colframe=boxcolor,colback=boxcolor]
\justify
\textbf{Abstract:}
JWST/NIRCam has recovered the stellar counterparts of many ALMA-detected, formerly HST-dark galaxies, yet ALMA has started identifying dusty star-forming galaxies (DSFGs) at $z>6$ that remain extremely faint or undetected even with NIRCam.
These ``NIRCam-dark'' DSFGs may trace a missing rapid-growth phase connecting UV-luminous galaxies at $z>10$ to massive quiescent galaxies at $z\gtrsim4$.
So far, the small, heterogeneously selected NIRCam-dark sample is insufficient to determine their abundance or role in the evolution of massive galaxies.
We show that a blind 1.3-mm continuum survey with ALMA2040 can provide the first high-significance census of this elusive NIRCam-dark DSFG population across $z\simeq 6 - 9$.
We conduct sensitivity and survey-speed calculations requiring a $10\sigma$ detection of a NIRCam-dark DSFG with a dust-obscured star-formation rate of $60\,M_\odot\,{\rm yr}^{-1}$ at $z = 8$.
At a like-for-like 7.5-GHz normalization, ALMA2040 will reach the required 1.3-mm RMS of $21\,\mu$Jy beam$^{-1}$ about 90 times faster than current ALMA.
With 100\,hr on-source, ALMA2040 can cover an area of $\sim0.5$\,deg$^2$ and detect $\langle N \rangle\sim16$ NIRCam-dark DSFGs over $7.5<z<9$, assuming a fiducial number density $n=3\times10^{-6}$ cMpc$^{-3}$.
Dividing the survey among independent fields with deep JWST coverage would mitigate cosmic variance and enable immediate NIRCam, MIRI, and spectroscopic characterization of their environments.
The $\geq512$-GHz instantaneous bandwidth expected for ALMA2040 will also detect [C~II] line at a wide redshift range, supplementing the continuum discovery with redshift confirmation.
Such a survey will test whether NIRCam-dark DSFGs trace a missing rapid-growth phase between the earliest UV-luminous galaxies at $z>10$ and the rapid emergence of massive quiescent systems at $z\gtrsim 4$.
\end{tcolorbox}

\section{Scientific context and open questions}

\subsection{From HST-dark to NIRCam-dark galaxies}

A key goal of (sub-)millimeter interferometry is to precisely localize submillimeter galaxies (SMGs), a population of distant, vigorously star-forming galaxies discovered with single-dish telescopes \citep{Smail1997}.
With interferometers such as ALMA, thousands of SMGs have been localized, and their counterparts from X-ray and UV/optical wavelengths to the radio have been identified and analyzed \citep[see a review by][]{hodge-dacunha2020}.

The long quest to understand HDF850.1 provides a clear example that the stellar counterparts of SMGs could be difficult to identify.
HDF850.1 was one of the first and brightest SMGs discovered in the Hubble Deep Field \citep{Hughes1998}.
It remained undetected at HST wavelengths even after the arrival of WFC3-IR \citep{Serjeant2014} and was therefore commonly regarded as a textbook example of an ``HST-dark'' galaxy.
Fourteen years after its initial discovery, a millimeter spectral scan with PdBI confirmed its redshift at $z=5.183$ through detections of CO and [C~II] emission \citep{Walter+12}.
This history demonstrates that (sub-)millimeter interferometry uniquely reveals massive galaxy formation that optical and near-infrared surveys cannot access alone.
HDF850.1 is not an exceptional observational curiosity.
Over the past decade, many ALMA observing programs have shown that ``HST-dark'' galaxies comprise approximately 15--20\% of dusty star-forming galaxies (DSFGs) found in blind continuum surveys \citep[e.g.,][]{Simpson2015,Franco+18,Wang+19}.

With JWST \citep{Gardner2023}, the stellar counterpart of HDF850.1 was finally detected and resolved through deep NIRCam observations (\citealt{Sun+24}; see also \citealt{Herard-Demanche+25}), revealing its spatial relation to the obscured starburst.
Similarly, JWST has revealed a large number of formerly HST-dark dusty star-forming systems, including AzTECC71, GN20, and MAMBO-9 \citep{2023ApJ...958...36S,McKinney23a,Colina2023,Akins+25}.
Together, JWST and ALMA now provide a clearer view of massive-galaxy evolution at $z \simeq 3 - 7$, with much better constraints on the stellar masses \citep[e.g.,][]{2026arXiv260630802X}, star-formation histories \citep[SFHs; e.g.,][]{2025ApJ...982..125C}, and environments \citep[e.g.,][]{Sun+24,2026A&A...705A.214L} of these early massive starbursts, as well as on their contribution to the cosmic star-formation history \citep[e.g.,][]{2025ApJ...980...12S,2026MNRAS.545f1961L}.
The high spatial resolution offered by the two facilities also resolves mergers \citep[e.g.,][]{2024ApJ...973...25K,Akins+25}, spirals and bars \citep[e.g.,][]{2023ApJ...942L...1W,2025PASJ...77..432U,2026ApJ..1006L..29B}, compact starbursts \citep[e.g.,][]{2023A&A...677A..34G,2025ApJ...978..165H}, patchy dust--stellar geometries \citep[e.g.,][]{Sun+24,2026arXiv260413408L}, and spatially inhomogeneous star-formation histories \citep[e.g.,][]{2024A&A...688A..53L} in these galaxies.
The goal of this white paper is not to review these outstanding achievements from joint JWST--ALMA analyses of DSFGs over the past four years.
Instead, we ask ourselves the following question to ensure that we have not missed something rather important: \textbf{has JWST now detected all dust-obscured star-forming galaxies across cosmic history?}

The history of infrared astronomy gives us strong reason to \textbf{doubt} that the census is complete.
Each advance from Spitzer/IRAC \citep[e.g.,][]{Yan+04,Daddi+09} to HST/WFC3-IR \citep[e.g.,][]{Huang+11,Wang+16} and JWST/NIRCam \citep[e.g.,][]{Barrufet+23,Williams+24} has revealed dust-obscured star-forming galaxies at earlier epochs than previously expected.
ALMA had already identified dust-obscured galaxies at $z\sim7$ that remained undetected in deep HST and Spitzer/IRAC imaging \citep{Fudamoto2021}.
Therefore, the search for ``NIRCam-dark'' galaxies, i.e., galaxies that remain undetected even in deep NIRCam imaging, has become a natural science goal for joint ALMA and JWST continuum imaging surveys.
\citet{Perez-Gonzalez+24} presented the discovery of Cerberus, a NIRCam-dark galaxy in the Hubble Ultra Deep Field (HUDF) that is  detected only with MIRI at 10\,\micron\ and remains undetected in individual NIRCam bands down to $\sim30.5$\,AB mag ($5\sigma$).
The redshift and physical nature of Cerberus remain uncertain to date.
However, the ALMA 1.2\,mm non-detection with the deep ASPECS data \citep[][RMS of $9.3\,\mu$Jy beam$^{-1}$ over 4.2\,arcmin$^2$]{gonzalez-lopez2020}  disfavors a DSFG interpretation similar to those HST-dark galaxies mentioned before.
There have also been other NIRCam-dark ALMA-only sources that have been reported by blank-field surveys \citep[e.g.,][]{fujimoto2024-dualz} or as [C~II] emitters in quasar fields \citep[e.g.,][]{bakx2024}.
Most of these have modest significance (S/N $\sim5$ for  continuum and $\sim6$ for emission lines) and thus require confirmation through further observations.

\begin{figure}[!t]
\centering
\includegraphics[width=\linewidth]{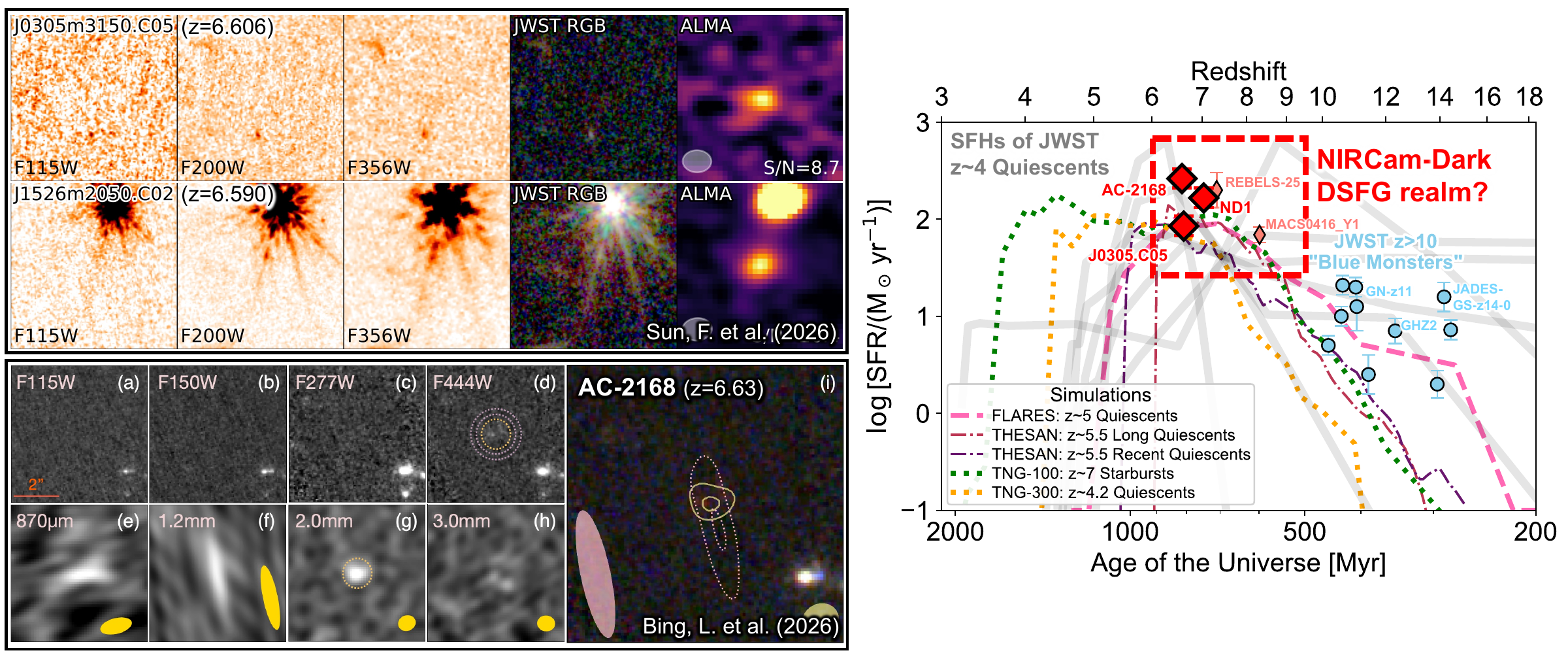}
\caption{Examples of NIRCam-dark DSFGs discovered through joint ALMA and JWST surveys.
\textit{Top-left:} Two NIRCam-dark DSFGs identified as quasar companions through the ASPIRE survey (adapted from \citealt{Sun2026}).
\textit{Bottom-left:} The NIRCam-dark DSFG AC-2168 discovered through A$^3$COSMOS (adapted from \citealt{Bing2026}).
\textit{Right:} The inferred star-formation histories of quiescent galaxies at $z\sim4$ \citep[e.g.,][]{carnall2023,glazebrook2024,baker2025} through JWST spectroscopy (gray curves) as compared with various simulations \citep[e.g.,][]{lovell2023,hartley2023,chittenden2025}.
For comparison, selected JWST-discovered blue monsters at $z>10$ are shown \citep[e.g.,][]{Bunker2023,Carniani2024,Castellano2024,Naidu2025}.
Three recently identified NIRCam-dark galaxies at $z\sim7$ are shown as red diamonds, and we highlight the redshift and SFR range in which NIRCam-dark DSFGs may reside (figure adapted from \citealt{Sun2026}).
}
\label{fig:nrc-dark}
\end{figure}

\subsection{An emerging NIRCam-dark dusty star-forming galaxy population at $z>6$}

A few particularly interesting NIRCam-dark DSFGs have been identified at $z > 6$ through ALMA continuum imaging and, again, [C~II] spectroscopy.
Through the ALMA ASPIRE large program \citep{Wang-feige-2026}, \citet{Sun2026} identified two NIRCam-dark DSFGs as companions to quasars at $z = 6.6$ (Figure~\ref{fig:nrc-dark}).
With ALMA 1.2-mm flux densities of 0.3--1.0\,mJy, these sources remain very faint in the NIRCam F356W band (down to $>28.0$\,mag), while their inferred star-formation rates (SFRs) are approximately 80--250\,$M_\odot\,{\rm yr}^{-1}$.
The redshifts of these galaxies are precisely determined by the secure detection of [C~II] and CO lines.
\citet{Sun2026} define the ``NIRCam-dark'' galaxy class using the observed-frame color:
\begin{equation}
   S_{3.6\,\mu \mathrm{m}}/S_\mathrm{1.2\,mm} < 10^{-4}
\end{equation}

As this color decreases with increasing redshift, it can effectively select DSFGs that are at higher redshifts and more dust-obscured than the majority of known DSFGs.
We emphasize that this color selection does not strictly require a non-detection in NIRCam imaging, which depends on its wavelength coverage and depth, as well as the assumed source size.
Because the two NIRCam-dark galaxies are identified through ASPIRE, a biased quasar survey, it is difficult to constrain the number density of such a NIRCam-dark population near $z \sim 7$.
\citet{Sun2026} assume a dark matter halo occupation model for both quasars and NIRCam-dark galaxies, and infer a number density of $n=3^{+4}_{-2}\times10^{-6}$ cMpc$^{-3}$ if NIRCam-dark galaxies can exist in relatively light halos.
However, if NIRCam-dark galaxies are associated only with the most massive halos, they would remain as rare as $n \sim 10^{-8}$\,cMpc$^{-3}$.
Independent assessment of the abundance of NIRCam-dark galaxies will need to be obtained from joint ALMA and JWST surveys in blank fields.

Most recently, independent discoveries show that these systems are not confined to the massive halos in the Epoch of Reionization anchored by quasars.
\citet{Bing2026} reported the blind discovery of AC-2168 through ALMA 2\,mm data in the COSMOS field ($0.20\pm0.02$\,mJy), and confirmed it at $z=6.631$ through [C~II] ($1.2\pm0.2$\,Jy\,km\,s$^{-1}$) and tentative CO line detections (Figure~\ref{fig:nrc-dark}).
It satisfies the aforementioned NIRCam-dark criterion with a non-detection in the F277W band ($>28.0$\,mag) and only $\sim26.2$\,mag in the F444W band.
AC-2168 has ${\rm SFR}=244\,M_\odot\,{\rm yr}^{-1}$ and $M_\star=3.7\times10^{10}\,M_\odot$ \citep{Bing2026}, comparable to those in \citet{Sun2026}.
A second example, ND1, was discovered as a NOEMA millimeter continuum source in the GOODS-N field, near the sightline to the red quasar GNz7q \citep{Fujimoto+22}.
ND1 is undetected in all NIRCam short-wavelength bands, also satisfying the aforementioned NIRCam-dark criterion.
\citet{2026arXiv260917680F} reported secure JWST/NIRSpec detections of [O~III] and H$\alpha$ line emission, together with a $\sim5\sigma$ NOEMA [C~II] detection ($0.82\pm0.30$\,Jy\,km\,s$^{-1}$), establishing ND1 as a DSFG with an SFR of $168\pm35\,M_\odot\,{\rm yr}^{-1}$ at $z = 7.03$.
Together, these discoveries suggest that NIRCam-dark galaxies may be of reasonable abundance in the Epoch of Reionization (e.g., $n = 7.8_{-6.5}^{+18.0}\times10^{-6}$\,cMpc$^{-3}$ from \citealt{Bing2026}).
This broadly agrees with the fiducial $3\times10^{-6}$\,cMpc$^{-3}$ light-halo estimate from \citet{Sun2026}.
However, a proper number-density measurement is still very challenging given the small and heterogeneous sample collected so far.

\subsection{The missing rapid-growth phase, anchored by NIRCam-dark galaxies}

NIRCam-dark starbursts may represent a missing rapid-growth phase between the unexpectedly luminous ``blue monsters'' at $z>10$ and massive quiescent galaxies already in place by $z\gtrsim4$ \citep[Figure~\ref{fig:nrc-dark}, right panel;][]{Sun2026}.
Both populations have posed significant challenges to theories of massive galaxy formation and evolution.
On the highest-redshift side, JWST has spectroscopically confirmed luminous galaxies out to $z > 14$ \citep[e.g., JADES-GS-z14-0;][]{Carniani2024}.
Under halo-growth models with a fixed baryon-to-star conversion efficiency, such galaxies should evolve into galaxies with ${\rm SFR}\gtrsim100\,M_\odot\,{\rm yr}^{-1}$ by $z\sim9$, which would appear as UV-luminous ($M_{\rm UV}<-23$\,mag) sources if their star formation were unobscured \citep{Sun2026}.
However, such a luminous population has not yet been found in the deep NIRCam surveys that discovered these blue monsters, and the closest examples at $z\sim9$ are at $M_{\rm UV}\sim-22$\,mag (e.g., Gz9p3, \citealt{Boyett2023}; EGSY8p7, \citealt{Zitrin2015}).
On the lower-redshift side, massive quiescent galaxies at $3<z<5$ have been routinely identified through the same NIRCam surveys and thus in comparable volumes, and their inferred star-formation histories require that many of them underwent major starbursts with ${\rm SFR}\gtrsim100\,M_\odot\,{\rm yr}^{-1}$ at $z\simeq6 - 9$ \citep[e.g.,][]{Carnall+24}.
Again, such starburst progenitors remain largely unidentified in the NIRCam surveys conducted so far.
In other words, we may have missed a substantial population of the most massive galaxies in their most rapid growth phase at $z\simeq6 - 9$, as inferred from both directions.
This could simply be explained by a heavily dust-obscured phase that is therefore invisible to NIRCam.

These NIRCam-dark, and potentially ``JWST-dark,'' galaxies may be among the first systems in which more than 90\% of ongoing star formation is hidden at rest-frame UV wavelengths \citep{Sun2026}.
Patchy dust--stellar geometries may leave a blue but intrinsically faint rest-UV component (e.g., MACS0416-Y1 at $z=8.31$; \citealt{2023ApJ...952....9T,2024ApJ...977L..36H}).
Such a compact UV-emitting region can remain NIRCam-faint and spatially decoupled from the dominant dust continuum, as in HDF850.1, and therefore does not trace the total SFR of the galaxy.
Based on the aforementioned SFRs and inferred number densities of NIRCam-dark galaxies, existing ALMA blank-field surveys such as ASPECS \citep{gonzalez-lopez2020}, ASAGAO \citep{2018PASJ...70..105H}, GOODS-ALMA \citep{gomez-guijarro2022}, MORA \citep{Casey+21}, and CHAMPS \citep{2026arXiv260935992F} do not reach the required depth and area simultaneously (Figure~\ref{fig:yield}, panel a), and therefore miss these rare systems or remain limited by cosmic variance and shot noise.
Therefore, without a proper blank-field ALMA continuum survey program, we do not yet know the number density and thus the role of NIRCam-dark DSFGs in the evolution of massive galaxies.

\smallskip

Below we summarize a few open questions for the investigation of NIRCam-dark DSFGs:
\begin{compactenum}
\item What is the true number density of NIRCam-dark starbursts at $z\simeq6 - 9$, and how rapidly do they evolve and assemble stellar mass as a function of time?
\item What is the fraction of cosmic stellar mass density and cosmic star formation hidden from even the deepest NIRCam surveys at this epoch?
\item Are these systems common progenitors of early quiescent galaxies at $z \gtrsim 4$ (i.e., through evidence such as matched abundance, size consistency, and similarity in clustering properties)?
\item Are their star-formation efficiency and baryon-to-stellar-mass conversion efficiency consistent with those of blue monsters at $z \gtrsim 10$?
\end{compactenum}
Deep, blank-field surveys with ALMA2040 would deliver new insights into these open questions, providing a statistical characterization of NIRCam-dark galaxies as an emerging population.

\section{ALMA2040: Technical Requirements}

\subsection{1.3-mm blank-field detection experiment}

We consider the detectability of NIRCam-dark DSFGs at $z \sim 8$ through a blind 1.3-mm (230\,GHz) continuum survey with ALMA2040.
At $z\simeq6$--9, 1.3\,mm samples the long-wavelength side of the dust SED with a favorable negative $K$-correction, while remaining close to the redshifted [C~II] line.
Compared with other ALMA wavelengths, $\sim1.3$\,mm offers a balance among continuum depth, survey speed, and redshift information.
Following \citet{Sun2026}, we adopt the best-fit SED of J0305m3150.C05, shift it to $z=8$, and scale it to ${\rm SFR}=60\,M_\odot\,{\rm yr}^{-1}$.
We acknowledge uncertainties in the far-IR SED shape and the conversion from luminosity to SFR.
Therefore, the calculation presented here serves only as a proof of concept, while the far-IR SED shape will be better characterized by the wide bandwidth offered by ALMA2040.
The predicted continuum flux density at 230 GHz is $S_\mathrm{230\,GHz}=0.21$ mJy.
We require $\geq10\sigma$ detections for secure identification and size measurements, which set an RMS depth of $\sigma_\mathrm{230\,GHz}=21\,\mu$Jy beam$^{-1}$.
We adopt the point-source sensitivities expected for ALMA2040 at a Decl.\ of $-23^{\circ}$.

For a 12-m half-power beam width (HPBW) of $25''$ at 230\,GHz, the squared primary-beam response ($\mathrm{PB}^2$) gives a noise-weighted field of view of $\sim0.10$\,arcmin$^2$ per pointing.
Assuming a flat cosmology with $H_0=70$ km s$^{-1}$ Mpc$^{-1}$ and $\Omega_m=0.3$, the comoving volume over $7.5<z<9.0$ is then $dV / d\Omega = 2866$\,cMpc$^3$\,arcmin$^{-2}$.
The area for one expected detection of a NIRCam-dark DSFG ($\langle N\rangle=1$) is
\begin{equation}
 A_{\langle N\rangle=1}=\left[n\,(dV/d\Omega)\right]^{-1}.
\end{equation}

For a like-for-like 7.5-GHz bandwidth, the adopted one-hour sensitivities are 14.7\,$\mu$Jy beam$^{-1}$ for current ALMA and 1.56\,$\mu$Jy beam$^{-1}$ for ALMA2040.
Because mapping speed scales as $\sigma^{-2}$, ALMA2040 is $\sim90$ times faster for this detection experiment.
We note that our sensitivity calculation is only for on-source time, and the real observing implementation must also consider multiple overheads including calibration and slewing.
We also note that current ALMA2040 concept targets at least 512\,GHz of instantaneous bandwidth, which will deliver much faster surveys, emission-line scans, and dust-temperature measurements.
This calculation is not intended to provide a detailed design for an ALMA2040 observing program.

For a number density of $n=3\times10^{-6}$ cMpc$^{-3}$ derived by \citet{Sun2026} assuming a light halo occupation model, it will require a survey area of $\sim$116 arcmin$^2$ to detect one NIRCam-dark DSFG in the aforementioned redshift range.
This is equivalent to $\sim$half of the JADES survey area in GOODS-S/N fields \citep{eisenstein2026} with a coverage of at least eight NIRCam filters.
ALMA2040 will complete a survey of this area at the required depth in about 6.4\,hr.
In contrast, the current ALMA facility would require 565\,hr to detect one such source for the same survey area and depth.
Therefore, while detecting an individual bright NIRCam-dark galaxy with the current ALMA is possible through archival studies (e.g., \citealt{Bing2026} with A$^{3}$COSMOS), characterizing this population, especially their number density, remains prohibitively expensive.
However, an ALMA2040 program of $\sim100$\,hr will likely move beyond a handful of NIRCam-dark DSFG detections and begin to deliver population-wide statistics.

\begin{figure}[!t]
\centering
\includegraphics[width=\textwidth]{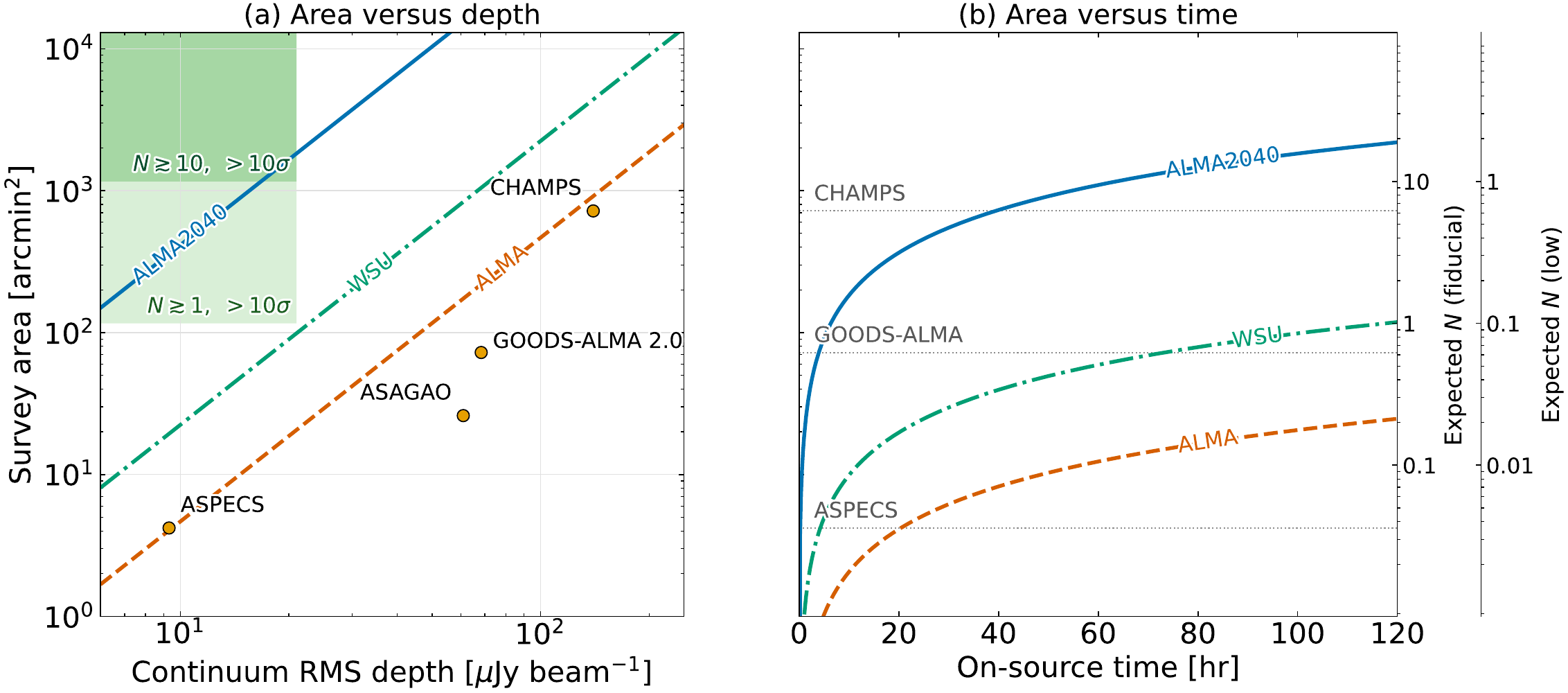}
\caption{Expected survey yield of NIRCam-dark DSFGs over $7.5<z<9$.
Left: survey area versus continuum RMS depth for 100-hr on-source surveys with current ALMA, ALMA WSU, and ALMA2040.
ASPECS, ASAGAO, GOODS-ALMA, and CHAMPS are shown for comparison \citep{gonzalez-lopez2020,2018PASJ...70..105H,gomez-guijarro2022,2026arXiv260935992F}.
The shaded regions mark surveys reaching $N\gtrsim1$ and $N\gtrsim10$ detections at $\geq10\sigma$ for the fiducial source flux density (0.21\,mJy) and number density ($n=3\times10^{-6}$\,cMpc$^{-3}$).
Right: survey area at RMS\,=\,21\,$\mu$Jy\,beam$^{-1}$ as a function of on-source time for current ALMA, ALMA WSU, and ALMA2040.
The right axes give the expected yields for the fiducial and low ($n=3\times10^{-7}$\,cMpc$^{-3}$) number densities.
Horizontal dotted lines mark the survey area of the ASPECS, GOODS-ALMA, and CHAMPS.
}
\label{fig:yield}
\end{figure}

Figure~\ref{fig:yield} compares the survey speed of current ALMA, the WSU, and ALMA2040, and places a few current blank-field surveys in the area--depth plane.
Although the WSU will improve the Band-6 imaging speed by a factor of 4.8, it remains insufficient for a population census within a practical survey time \citep{Carpenter2022,ALMAWSU2026}.
In addition, AtLAST will detect such sources rapidly over much wider fields \citep{2026arXiv260705022C}, but its $\sim6\farcs7$ beam will make blending and counterpart identification major challenges.
In contrast, ALMA2040 will uniquely combine the required survey speed with sub-arcsec imaging and wideband spectroscopy for source identification and redshift confirmation.

\begin{table}[ht]
\centering
\caption{Expected yield of NIRCam-dark DSFGs from a 100-hr (on-source) blind ALMA 1.3-mm continuum survey down to RMS\,$=$\,21\,$\mu$Jy beam$^{-1}$. Two number densities are assumed, and $P_{\rm low}(\geq1)$ is the chance of obtaining at least one detection assuming the low density ($n = 3\times10^{-7}$\,cMpc$^{-3}$).}
\label{tab:survey}
\small
\begin{tabular}{lcccc}
\toprule
Facility & Area & $\langle N\rangle_{3\times10^{-6}}$ & $\langle N\rangle_{3\times10^{-7}}$ & $P_{\rm low}(\geq1)$ \\
 & (arcmin$^2$) & & & \\
\midrule
Current ALMA & 20.6 & 0.18 & 0.018 & 1.8\% \\
ALMA WSU & 98.9 & 0.85 & 0.085 & 8.2\% \\
ALMA2040 & 1830 & 15.7 & 1.57 & 79\% \\
\bottomrule
\end{tabular}
\end{table}

A 100-hr ALMA2040 survey at 1.3\,mm covers approximately 1830~arcmin$^2$ ($0.51$~deg$^2$).
Over $7.5<z<9$, this area corresponds to a comoving volume of $5.24\times10^6$ cMpc$^3$.
We expect about 16 NIRCam-dark galaxies at the fiducial density.
If the abundance is ten times lower, we expect to detect 1.6 sources, with a 79\% probability of at least one detection (Table~\ref{tab:survey}).
With current ALMA, a program of the same on-source time would cover only 20.6\,arcmin$^2$, making it unlikely to detect any NIRCam-dark DSFG in the low-density case ($P_{\rm low}(\geq1) = 1.8\%$).
The WSU would cover $\sim99$\,arcmin$^2$ in 100 hours and yield $\langle N\rangle\simeq0.85$ at the fiducial density, still below the requirement for a population-wide study.

We caution about the impact of cosmic variance when interpreting the observed counts from such a blind continuum survey.
For example, \citet{Zavala2026} estimate an approximately 13\% cosmic sample variance over the 0.18-deg$^2$ area observed by CHAMPS at $z = 6 - 8$.
The exact cosmic variance of NIRCam-dark galaxies depends on their unknown halo bias, and we adopt a plausible 10--20\% contribution for the 100-hr survey.
Although this is below the Poisson uncertainties of the source counts in both cases,
dividing the total survey area among several independent, well-studied extragalactic fields would further reduce the impact of cosmic variance and provide immediate multiwavelength coverage from a large number of surveys.

Additionally, we highlight the opportunity to conduct a [C~II] redshift scan simultaneously with the continuum survey.
The [C~II] 158\,$\mu$m line will span 223.6 to 190.1\,GHz ($\Delta \nu = 33.5$\,GHz) over the targeted redshift range of $7.5<z<9$, and 190.0--271.5\,GHz over $6<z<9$.
ALMA2040 targets for an instantaneous bandwidth at $\geq512$\,GHz, and may therefore cover this full interval, while multiple CO lines can identify lower-redshift interlopers.
For ${\rm SFR}=60\,M_\odot\,{\rm yr}^{-1}$, the empirical SFR--$L_{\rm [CII]}$ relation implies $L_{\rm [CII]}\sim6\times10^8\,L_\odot$ and $S\Delta v\sim0.42$\,Jy\,km\,s$^{-1}$ at $z=8$ \citep{DeLooze2014}.
For a 300\,km\,s$^{-1}$ line, ALMA2040 will reach an integrated line RMS of $\sim0.036$\,Jy\,km\,s$^{-1}$ during the continuum-survey dwell, leading to a [C\,II] line detection near $12\sigma$ and thus a redshift confirmation.
The broad bandwidth of ALMA2040 will facilitate [C\,II] and [O\,III] 88\,$\mu$m spectroscopic follow-up even if the line emission is weak, for example, owing to scatter in the SFR--$L_{\rm [CII]}$ relation.

\section{Synergies with current and future facilities}

The scientific return of the proposed blind continuum survey will be maximized in extragalactic fields with deep JWST coverage.
Deep NIRCam imaging is required to establish the NIRCam-dark nature of the detected sources and thus place them at high redshifts.
At the same time, deep MIRI imaging samples their rest-frame optical emission at $z\gtrsim6$ and thus constrains their stellar masses and dust attenuation.
Additionally, NIRCam WFSS and NIRSpec spectroscopy can efficiently characterize the environments of NIRCam-dark DSFGs, in which UV-bright and less dusty companion galaxies may be identified \citep[e.g., through ASPIRE-like surveys;][]{Wang-feige-2026}.
Finally, the upcoming SKA1-Mid ultra-deep surveys may also detect NIRCam-dark DSFGs at $z\gtrsim6$ \citep{Wagg2014}.
Together, these synergies will transform a blind millimeter continuum census into a complete physical characterization of the NIRCam-dark DSFG population.

\section*{Acknowledgements}
The authors thank Francesca Rizzo for leading and coordinating the efforts within the ALMA2040 high-z working group.
The authors thank Mar\'ia Teresa Beltr\'an Sorolla for her careful and constructive review.
This work was developed as part of the broader ALMA2040 community effort.
We acknowledge the Lorentz Center and NOVA for supporting the community workshops and coordination activities that helped initiate and shape the ALMA2040 White Paper process.
The authors acknowledge the use of generative artificial intelligence (GPT 5.6 Sol and Claude Fable 5) in preparing this manuscript, under human supervision and review.
The authors are solely responsible for all content presented in this manuscript.

\medskip

\enlargethispage{1\baselineskip}
\bibliographystyle{apj_mod}
{\bibliography{main}}

@article{Hughes1998,
  author =        {{Hughes}, David H. and {Serjeant}, Stephen and
                   {Dunlop}, James and {Rowan-Robinson}, Michael and
                   {Blain}, Andrew and {Mann}, Robert G. and
                   {Ivison}, Rob and {Peacock}, John and
                   {Efstathiou}, Andreas and {Gear}, Walter and
                   {Oliver}, Seb and {Lawrence}, Andy and
                   {Longair}, Malcolm and {Goldschmidt}, Pippa and
                   {Jenness}, Tim},
  journal =       {\nat},
  month =         jul,
  number =        {6690},
  pages =         {241-247},
  title =         {{High-redshift star formation in the Hubble Deep
                   Field revealed by a submillimetre-wavelength survey}},
  volume =        {394},
  year =          {1998},
  doi =           {10.1038/28328},
}

@article{Carniani2024,
  author =        {{Carniani}, Stefano and {Hainline}, Kevin and
                   {D'Eugenio}, Francesco and {Eisenstein}, Daniel J. and
                   {Jakobsen}, Peter and {Witstok}, Joris and
                   {Johnson}, Benjamin D. and {Chevallard}, Jacopo and
                   {Maiolino}, Roberto and {Helton}, Jakob M. and
                   {Willott}, Chris and {Robertson}, Brant and
                   {Alberts}, Stacey and {Arribas}, Santiago and
                   {Baker}, William M. and {Bhatawdekar}, Rachana and
                   {Boyett}, Kristan and {Bunker}, Andrew J. and
                   {Cameron}, Alex J. and {Cargile}, Phillip A. and
                   {Charlot}, St{\'e}phane and {Curti}, Mirko and
                   {Curtis-Lake}, Emma and {Egami}, Eiichi and
                   {Giardino}, Giovanna and {Isaak}, Kate and
                   {Ji}, Zhiyuan and {Jones}, Gareth C. and
                   {Kumari}, Nimisha and {Maseda}, Michael V. and
                   {Parlanti}, Eleonora and
                   {P{\'e}rez-Gonz{\'a}lez}, Pablo G. and {Rawle}, Tim and
                   {Rieke}, George and {Rieke}, Marcia and
                   {Del Pino}, Bruno Rodr{\'\i}guez and {Saxena}, Aayush and
                   {Scholtz}, Jan and {Smit}, Renske and {Sun}, Fengwu and
                   {Tacchella}, Sandro and {{\"U}bler}, Hannah and
                   {Venturi}, Giacomo and {Williams}, Christina C. and
                   {Willmer}, Christopher N.~A.},
  journal =       {\nat},
  month =         sep,
  number =        {8029},
  pages =         {318-322},
  title =         {{Spectroscopic confirmation of two luminous galaxies
                   at a redshift of 14}},
  volume =        {633},
  year =          {2024},
  doi =           {10.1038/s41586-024-07860-9},
}

@article{Walter+12,
  author =        {{Walter}, Fabian and {Decarli}, Roberto and
                   {Carilli}, Chris and {Bertoldi}, Frank and
                   {Cox}, Pierre and {da Cunha}, Elisabete and
                   {Daddi}, Emanuele and {Dickinson}, Mark and
                   {Downes}, Dennis and {Elbaz}, David and
                   {Ellis}, Richard and {Hodge}, Jacqueline and
                   {Neri}, Roberto and {Riechers}, Dominik A. and
                   {Weiss}, Axel and {Bell}, Eric and
                   {Dannerbauer}, Helmut and {Krips}, Melanie and
                   {Krumholz}, Mark and {Lentati}, Lindley and
                   {Maiolino}, Roberto and {Menten}, Karl and
                   {Rix}, Hans-Walter and {Robertson}, Brant and
                   {Spinrad}, Hyron and {Stark}, Dan P. and
                   {Stern}, Daniel},
  journal =       {\nat},
  month =         jun,
  number =        {7402},
  pages =         {233-236},
  title =         {{The intense starburst HDF{\,}850.1 in a galaxy
                   overdensity at z{\,}{\ensuremath{\approx}}{\,}5.2 in
                   the Hubble Deep Field}},
  volume =        {486},
  year =          {2012},
  doi =           {10.1038/nature11073},
}

@article{Franco+18,
  author =        {{Franco}, M. and {Elbaz}, D. and {B{\'e}thermin}, M. and
                   {Magnelli}, B. and {Schreiber}, C. and {Ciesla}, L. and
                   {Dickinson}, M. and {Nagar}, N. and {Silverman}, J. and
                   {Daddi}, E. and {Alexander}, D.~M. and {Wang}, T. and
                   {Pannella}, M. and {Le Floc'h}, E. and {Pope}, A. and
                   {Giavalisco}, M. and {Maury}, A.~J. and
                   {Bournaud}, F. and {Chary}, R. and {Demarco}, R. and
                   {Ferguson}, H. and {Finkelstein}, S.~L. and
                   {Inami}, H. and {Iono}, D. and {Juneau}, S. and
                   {Lagache}, G. and {Leiton}, R. and {Lin}, L. and
                   {Magdis}, G. and {Messias}, H. and {Motohara}, K. and
                   {Mullaney}, J. and {Okumura}, K. and {Papovich}, C. and
                   {Pforr}, J. and {Rujopakarn}, W. and {Sargent}, M. and
                   {Shu}, X. and {Zhou}, L.},
  journal =       {\aap},
  month =         dec,
  pages =         {A152},
  title =         {{GOODS-ALMA: 1.1 mm galaxy survey. I. Source catalog
                   and optically dark galaxies}},
  volume =        {620},
  year =          {2018},
  doi =           {10.1051/0004-6361/201832928},
  eid =           {A152},
}

@article{Wang+19,
  author =        {{Wang}, T. and {Schreiber}, C. and {Elbaz}, D. and
                   {Yoshimura}, Y. and {Kohno}, K. and {Shu}, X. and
                   {Yamaguchi}, Y. and {Pannella}, M. and {Franco}, M. and
                   {Huang}, J. and {Lim}, C. -F. and {Wang}, W. -H.},
  journal =       {\nat},
  month =         aug,
  number =        {7768},
  pages =         {211-214},
  title =         {{A dominant population of optically invisible massive
                   galaxies in the early Universe}},
  volume =        {572},
  year =          {2019},
  doi =           {10.1038/s41586-019-1452-4},
}

@article{Casey+21,
  author =        {{Casey}, Caitlin M. and {Zavala}, Jorge A. and
                   {Manning}, Sinclaire M. and {Aravena}, Manuel and
                   {B{\'e}thermin}, Matthieu and {Caputi}, Karina I. and
                   {Champagne}, Jaclyn B. and {Clements}, David L. and
                   {Drew}, Patrick and {Finkelstein}, Steven L. and
                   {Fujimoto}, Seiji and {Hayward}, Christopher C. and
                   {Dekel}, Anton M. and {Kokorev}, Vasily and
                   {Lagos}, Claudia del P. and {Long}, Arianna S. and
                   {Magdis}, Georgios E. and {Man}, Allison W.~S. and
                   {Mitsuhashi}, Ikki and {Popping}, Gerg{\"o} and
                   {Spilker}, Justin and {Staguhn}, Johannes and
                   {Talia}, Margherita and {Toft}, Sune and
                   {Treister}, Ezequiel and {Weaver}, John R. and
                   {Yun}, Min},
  journal =       {\apj},
  month =         dec,
  number =        {2},
  pages =         {215},
  title =         {{Mapping Obscuration to Reionization with ALMA
                   (MORA): 2 mm Efficiently Selects the Highest-redshift
                   Obscured Galaxies}},
  volume =        {923},
  year =          {2021},
  doi =           {10.3847/1538-4357/ac2eb4},
  eid =           {215},
}

@article{Fujimoto+22,
  author =        {{Fujimoto}, S. and {Brammer}, G.~B. and {Watson}, D. and
                   {Magdis}, G.~E. and {Kokorev}, V. and {Greve}, T.~R. and
                   {Toft}, S. and {Walter}, F. and {Valiante}, R. and
                   {Ginolfi}, M. and {Schneider}, R. and {Valentino}, F. and
                   {Colina}, L. and {Vestergaard}, M. and
                   {Marques-Chaves}, R. and {Fynbo}, J.~P.~U. and
                   {Krips}, M. and {Steinhardt}, C.~L. and {Cortzen}, I. and
                   {Rizzo}, F. and {Oesch}, P.~A.},
  journal =       {\nat},
  month =         apr,
  number =        {7905},
  pages =         {261-265},
  title =         {{A dusty compact object bridging galaxies and quasars
                   at cosmic dawn}},
  volume =        {604},
  year =          {2022},
  doi =           {10.1038/s41586-022-04454-1},
}

@article{Sun+24,
  author =        {{Sun}, Fengwu and {Helton}, Jakob M. and
                   {Egami}, Eiichi and {Hainline}, Kevin N. and
                   {Rieke}, George H. and {Willmer}, Christopher N.~A. and
                   {Eisenstein}, Daniel J. and {Johnson}, Benjamin D. and
                   {Rieke}, Marcia J. and {Robertson}, Brant and
                   {Tacchella}, Sandro and {Alberts}, Stacey and
                   {Baker}, William M. and {Bhatawdekar}, Rachana and
                   {Boyett}, Kristan and {Bunker}, Andrew J. and
                   {Charlot}, Stephane and {Chen}, Zuyi and
                   {Chevallard}, Jacopo and {Curtis-Lake}, Emma and
                   {Danhaive}, A. Lola and {DeCoursey}, Christa and
                   {Ji}, Zhiyuan and {Lyu}, Jianwei and
                   {Maiolino}, Roberto and {Rujopakarn}, Wiphu and
                   {Sandles}, Lester and {Shivaei}, Irene and
                   {{\"U}bler}, Hannah and {Willott}, Chris and
                   {Witstok}, Joris},
  journal =       {\apj},
  month =         jan,
  number =        {1},
  pages =         {69},
  title =         {{JADES: Resolving the Stellar Component and
                   Filamentary Overdense Environment of Hubble Space
                   Telescope (HST)-dark Submillimeter Galaxy HDF850.1 at
                   z = 5.18}},
  volume =        {961},
  year =          {2024},
  doi =           {10.3847/1538-4357/ad07e3},
  eid =           {69},
}

@article{Perez-Gonzalez+24,
  author =        {{P{\'e}rez-Gonz{\'a}lez}, Pablo G. and
                   {Rinaldi}, Pierluigi and {Caputi}, Karina I. and
                   {{\'A}lvarez-M{\'a}rquez}, Javier and
                   {Annunziatella}, Marianna and {Langeroodi}, Danial and
                   {Moutard}, Thibaud and {Boogaard}, Leindert and
                   {Iani}, Edoardo and {Melinder}, Jens and
                   {Costantin}, Luca and {{\"O}stlin}, G{\"o}ran and
                   {Colina}, Luis and {Greve}, Thomas R. and
                   {Wright}, Gillian and {Alonso-Herrero}, Almudena and
                   {Bik}, Arjan and {Bosman}, Sarah E.~I. and
                   {Crespo G{\'o}mez}, Alejandro and {Dicken}, Daniel and
                   {Eckart}, Andreas and
                   {Garc{\'\i}a-Mar{\'\i}n}, Macarena and
                   {Gillman}, Steven and {G{\"u}del}, Manuel and
                   {Henning}, Thomas and {Hjorth}, Jens and
                   {Jermann}, Iris and {Labiano}, {\'A}lvaro and
                   {Meyer}, Romain A. and
                   {Pei{\ensuremath{\beta}}ker}, Florian and
                   {Pye}, John P. and {Ray}, Thomas P. and
                   {Tikkanen}, Tuomo and {Walter}, Fabian and
                   {van der Werf}, Paul P.},
  journal =       {\apjl},
  month =         jul,
  number =        {1},
  pages =         {L10},
  title =         {{A NIRCam-dark Galaxy Detected with the MIRI/F1000W
                   Filter in the MIDIS/JADES Hubble Ultra Deep Field}},
  volume =        {969},
  year =          {2024},
  doi =           {10.3847/2041-8213/ad517b},
  eid =           {L10},
}

@article{Carnall+24,
  author =        {{Carnall}, A.~C. and {Cullen}, F. and {McLure}, R.~J. and
                   {McLeod}, D.~J. and {Begley}, R. and {Donnan}, C.~T. and
                   {Dunlop}, J.~S. and {Shapley}, A.~E. and
                   {Rowlands}, K. and {Almaini}, O. and
                   {Arellano-C{\'o}rdova}, K.~Z. and {Barrufet}, L. and
                   {Cimatti}, A. and {Ellis}, R.~S. and {Grogin}, N.~A. and
                   {Hamadouche}, M.~L. and {Illingworth}, G.~D. and
                   {Koekemoer}, A.~M. and {Leung}, H. -H. and
                   {Lovell}, C.~C. and {P{\'e}rez-Gonz{\'a}lez}, P.~G. and
                   {Santini}, P. and {Stanton}, T.~M. and {Wild}, V.},
  journal =       {\mnras},
  month =         oct,
  number =        {1},
  pages =         {325-348},
  title =         {{The JWST EXCELS survey: too much, too young, too
                   fast? Ultra-massive quiescent galaxies at 3 < z < 5}},
  volume =        {534},
  year =          {2024},
  doi =           {10.1093/mnras/stae2092},
}

@article{Herard-Demanche+25,
  author =        {{Herard-Demanche}, Thomas and {Bouwens}, Rychard J. and
                   {Oesch}, Pascal A. and {Naidu}, Rohan P. and
                   {Decarli}, Roberto and {Nelson}, Erica J. and
                   {Brammer}, Gabriel and {Weibel}, Andrea and
                   {Xiao}, Mengyuan and {Stefanon}, Mauro and
                   {Walter}, Fabian and {Matthee}, Jorryt and
                   {Meyer}, Romain A. and {Wuyts}, Stijn and
                   {Reddy}, Naveen and {Rowland}, Lucie and
                   {van Leeuwen}, Ivana and {Haro}, Pablo Arrabal and
                   {Dannerbauer}, Helmut and {Shapley}, Alice E. and
                   {Chisholm}, John and {van Dokkum}, Pieter and
                   {Labbe}, Ivo and {Illingworth}, Garth and
                   {Schaerer}, Daniel and {Shivaei}, Irene},
  journal =       {\mnras},
  month =         feb,
  number =        {2},
  pages =         {788-808},
  title =         {{Mapping dusty galaxy growth at z > 5 with FRESCO:
                   detection of H{\ensuremath{\alpha}} in submm galaxy
                   HDF850.1 and the surrounding overdense structures}},
  volume =        {537},
  year =          {2025},
  doi =           {10.1093/mnras/staf030},
}

@article{Akins+25,
  author =        {{Akins}, Hollis B. and {Casey}, Caitlin M. and
                   {Champagne}, Jaclyn B. and {Cooper}, Olivia and
                   {Franco}, Maximilien and {Fujimoto}, Seiji and
                   {Knudsen}, Kirsten K. and {Koekemoer}, Anton M. and
                   {Long}, Arianna S. and {Man}, Allison and
                   {Manning}, Sinclaire M. and {McKinney}, Jed and
                   {Zavala}, Jorge and {Arrabal Haro}, Pablo and
                   {Dickinson}, Mark and {Kokorev}, Vasily and
                   {Taylor}, Anthony J.},
  journal =       {\apj},
  month =         aug,
  number =        {2},
  pages =         {229},
  title =         {{JWST+ALMA Reveal the Interstellar Medium Kinematics
                   and Stellar Structure of MAMBO-9, a Merging Pair of
                   DSFGs in an Overdense Environment at z = 5.85}},
  volume =        {1006},
  year =          {2026},
  doi =           {10.3847/1538-4357/ae8449},
  eid =           {229},
  archivePrefix = {arXiv},
  eprint =        {2508.06607},
  primaryClass =  {astro-ph.GA},
  adsurl =        {https://ui.adsabs.harvard.edu/abs/2026ApJ..1006..229A},
}

@ARTICLE{Colina2023,
       author = {{Colina}, L. and {Crespo G{\'o}mez}, A. and {{\'A}lvarez-M{\'a}rquez}, J. and {Bik}, A. and {Walter}, F. and {Boogaard}, L. and {Labiano}, A. and {Peissker}, F. and {P{\'e}rez-Gonz{\'a}lez}, P. and {{\"O}stlin}, G. and {Greve}, T.~R. and {N{\o}rgaard-Nielsen}, H.~U. and {Wright}, G. and {Alonso-Herrero}, A. and {Azollini}, R. and {Caputi}, K.~I. and {Dicken}, D. and {Garc{\'\i}a-Mar{\'\i}n}, M. and {Hjorth}, J. and {Ilbert}, O. and {Kendrew}, S. and {Pye}, J.~P. and {Tikkanen}, T. and {van der Werf}, P. and {Costantin}, L. and {Iani}, E. and {Gillman}, S. and {Jermann}, I. and {Langeroodi}, D. and {Moutard}, T. and {Rinaldi}, P. and {Topinka}, M. and {van Dishoeck}, E.~F. and {G{\"u}del}, M. and {Henning}, Th. and {Lagage}, P.~O. and {Ray}, T. and {Vandenbussche}, B.},
        title = {{Uncovering the stellar structure of the dusty star-forming galaxy GN20 at z = 4.055 with MIRI/JWST}},
      journal = {\aap},
         year = {2023},
        month = may,
       volume = {673},
          eid = {L6},
        pages = {L6},
          doi = {10.1051/0004-6361/202346535},
archivePrefix = {arXiv},
       eprint = {2304.13529},
 primaryClass = {astro-ph.GA},
       adsurl = {https://ui.adsabs.harvard.edu/abs/2023A&A...673L...6C},
}

@ARTICLE{Sun2026,
       author = {{Sun}, Fengwu and {Yang}, Jinyi and {Wang}, Feige and {Eisenstein}, Daniel J. and {Decarli}, Roberto and {Fan}, Xiaohui and {Rieke}, George H. and {Ba{\~n}ados}, Eduardo and {Bosman}, Sarah E.~I. and {Cai}, Zheng and {Champagne}, Jaclyn B. and {Colina}, Luis and {D'Eugenio}, Francesco and {Fudamoto}, Yoshinobu and {Li}, Mingyu and {Lin}, Xiaojing and {Liu}, Weizhe and {Lyu}, Jianwei and {Mazzucchelli}, Chiara and {Jin}, Xiangyu and {Jun}, Hyunsung D. and {Tee}, Wei Leong and {Wu}, Yunjing and {Zhang}, Huanian},
        title = {{The Identification of Two JWST/NIRCam-dark Starburst Galaxies at z = 6.6 with ALMA}},
      journal = {\apj},
         year = {2026},
        month = jun,
       volume = {1003},
       number = {2},
          eid = {206},
        pages = {206},
          doi = {10.3847/1538-4357/ae66e3},
archivePrefix = {arXiv},
       eprint = {2506.06418},
 primaryClass = {astro-ph.GA},
       adsurl = {https://ui.adsabs.harvard.edu/abs/2026ApJ..1003..206S},
}

@ARTICLE{Bing2026,
       author = {{Bing}, Longji and {Oliver}, Seb and {Xiao}, Mengyuan and {Lagache}, Guilaine and {Adscheid}, Sylvia and {Liu}, Daizhong and {Magnelli}, Benjamin and {Neri}, Roberto and {Dessauges-Zavadsky}, Miroslava and {Koekemoer}, Anton M. and {Franco}, Maximilien and {Jin}, Shuowen and {Cooper}, Olivia R. and {Faisst}, Andreas L. and {Casey}, Catilin M. and {Kartaltepe}, Jeyhan S. and {Akins}, Hollis and {Beelen}, Alexandre and {Elbaz}, David and {Gillman}, Steven and {Harish}, Santosh and {Long}, Arianna S. and {McCracken}, Henry Joy and {Oesch}, Pascal and {Paquereau}, Louise and {Ponthieu}, Nicolas and {Rhodes}, Jason and {Robertson}, Brant and {Sanders}, David B. and {Shuntov}, Marko and {Wilkins}, Stephen},
        title = {{An almost NIRCam-dark dusty star-forming galaxy at z = 6.63}},
      journal = {\mnras},
         year = {2026},
        month = jun,
       volume = {549},
       number = {2},
          eid = {stag846},
        pages = {stag846},
          doi = {10.1093/mnras/stag846},
archivePrefix = {arXiv},
       eprint = {2511.08672},
 primaryClass = {astro-ph.GA},
       adsurl = {https://ui.adsabs.harvard.edu/abs/2026MNRAS.549ag846B},
}

@ARTICLE{Zavala2026,
       author = {{Zavala}, Jorge A. and {Faisst}, Andreas L. and {Aravena}, Manuel and {Casey}, Caitlin M. and {Kartaltepe}, Jeyhan S. and {Martinez}, III, Felix and {Silverman}, John D. and {Toft}, Sune and {Treister}, Ezequiel and {Akins}, Hollis B. and {Algera}, Hiddo and {Barboza}, Karina and {Battisti}, Andrew J. and {Brammer}, Gabriel and {Cai}, Zheng and {Champagne}, Jaclyn and {Drakos}, Nicole E. and {Egami}, Eiichi and {Fan}, Xiaohui and {Franco}, Maximilien and {Fudamoto}, Yoshinobu and {Fujimoto}, Seiji and {Gillman}, Steven and {Gozaliasl}, Ghassem and {Harish}, Santosh and {Jin}, Xiangyu and {Kakiichi}, Koki and {Kakkad}, Darshan and {Koekemoer}, Anton M. and {Lin}, Ruqiu and {Liu}, Daizhong and {Long}, Arianna S. and {Magdis}, Georgios E. and {Manning}, Sinclaire and {Martin}, Crystal L. and {McKinney}, Jed and {Meyer}, Romain and {Rodighiero}, Giulia and {Salazar}, Victoria and {Sanders}, David B. and {Shuntov}, Marko and {Talia}, Margherita and {Tanaka}, Takumi S. and {Wang}, Feige and {Wang}, Wuji and {Wilkins}, Stephen M. and {Yang}, Jinyi and {Yun}, Min S. and {The Champs} and {Cosmos-Web Collaborations}},
        title = {{ALMA and JWST Identification of Faint Dusty Star-forming Galaxies up to z {\ensuremath{\sim}} 8 and Their Connection with Other Galaxy Populations}},
      journal = {\apjl},
         year = {2026},
        month = feb,
       volume = {998},
       number = {2},
          eid = {L36},
        pages = {L36},
          doi = {10.3847/2041-8213/ae382a},
archivePrefix = {arXiv},
       eprint = {2512.16215},
 primaryClass = {astro-ph.GA},
       adsurl = {https://ui.adsabs.harvard.edu/abs/2026ApJ...998L..36Z},
}

@ARTICLE{2026arXiv260917680F,
       author = {{Fei}, Qinyue and {Fujimoto}, Seiji and {Brammer}, Gabriel and {Christensen}, Lise and {Vestergaard}, Marianne and {Naidu}, Rohan P. and {Simcoe}, Robert A. and {Murray}, Norman and {Ho}, Luis C. and {Li}, Ruancun and {Bromm}, Volker and {{\'A}lvarez-M{\'a}rquez}, Javier and {Akins}, Hollis B. and {Asada}, Yoshihisa and {Di Stefano}, Simona and {Inayoshi}, Kohei and {Kokorev}, Vasily and {Matthee}, Jorryt and {Meyer}, Romain A. and {Polzin}, Ava and {Valentino}, Francesco and {Walter}, Fabian and {Neeleman}, Marcel and {Wu}, Yunjing and {Xiao}, Mengyuan},
        title = "{A dense, metal-rich absorber driven by a heavily dust-obscured galaxy: The direct evidence of CGM enrichment at $z\sim7$}",
      journal = {arXiv e-prints},
         year = 2026,
        month = sep,
          eid = {arXiv:2609.17680},
        pages = {arXiv:2609.17680},
          doi = {10.48550/arXiv.2609.17680},
archivePrefix = {arXiv},
       eprint = {2609.17680},
 primaryClass = {astro-ph.GA},
       adsurl = {https://ui.adsabs.harvard.edu/abs/2026arXiv260917680F},
}

@ARTICLE{Serjeant2014,
       author = {{Serjeant}, Stephen and {Marchetti}, Lucia},
        title = "{Constraints on the galaxy `main sequence' at z > 5: the stellar mass of HDF850.1}",
      journal = {\mnras},
         year = 2014,
        month = oct,
       volume = {443},
       number = {4},
        pages = {3118-3126},
          doi = {10.1093/mnras/stu1266},
archivePrefix = {arXiv},
       eprint = {1409.0558},
 primaryClass = {astro-ph.GA},
       adsurl = {https://ui.adsabs.harvard.edu/abs/2014MNRAS.443.3118S}
}

@ARTICLE{Simpson2015,
       author = {{Simpson}, J.~M. and {Smail}, Ian and {Swinbank}, A.~M. and {Almaini}, O. and {Blain}, A.~W. and {Bremer}, M.~N. and {Chapman}, S.~C. and {Chen}, Chian-Chou and {Conselice}, C. and {Coppin}, K.~E.~K. and {Danielson}, A.~L.~R. and {Dunlop}, J.~S. and {Edge}, A.~C. and {Farrah}, D. and {Geach}, J.~E. and {Hartley}, W.~G. and {Ivison}, R.~J. and {Karim}, A. and {Lani}, C. and {Ma}, C.-J. and {Meijerink}, R. and {Micha{\l}owski}, M.~J. and {Mortlock}, A. and {Scott}, D. and {Simpson}, C.~J. and {Spaans}, M. and {Thomson}, A.~P. and {van Kampen}, E. and {van der Werf}, P.~P.},
        title = "{The SCUBA-2 Cosmology Legacy Survey: ALMA Resolves the Rest-frame Far-infrared Emission of Sub-millimeter Galaxies}",
      journal = {\apj},
         year = 2015,
        month = jan,
       volume = {799},
       number = {1},
          eid = {81},
        pages = {81},
          doi = {10.1088/0004-637X/799/1/81},
archivePrefix = {arXiv},
       eprint = {1411.5025},
 primaryClass = {astro-ph.GA},
       adsurl = {https://ui.adsabs.harvard.edu/abs/2015ApJ...799...81S}
}

@ARTICLE{Smail1997,
       author = {{Smail}, Ian and {Ivison}, R.~J. and {Blain}, A.~W.},
        title = "{A Deep Sub-millimeter Survey of Lensing Clusters: A New Window on Galaxy Formation and Evolution}",
      journal = {\apjl},
         year = 1997,
        month = nov,
       volume = {490},
       number = {1},
        pages = {L5-L8},
          doi = {10.1086/311017},
archivePrefix = {arXiv},
       eprint = {astro-ph/9708135},
       adsurl = {https://ui.adsabs.harvard.edu/abs/1997ApJ...490L...5S}
}

@ARTICLE{hodge-dacunha2020,
       author = {{Hodge}, Jacqueline A. and {da Cunha}, Elisabete},
        title = "{High-redshift Star Formation in the Atacama Large Millimetre/Submillimetre Array Era}",
      journal = {Royal Society Open Science},
         year = 2020,
        month = dec,
       volume = {7},
       number = {12},
          eid = {200556},
        pages = {200556},
          doi = {10.1098/rsos.200556},
archivePrefix = {arXiv},
       eprint = {2004.00934},
 primaryClass = {astro-ph.GA},
       adsurl = {https://ui.adsabs.harvard.edu/abs/2020RSOS....700556H}
}

@ARTICLE{McKinney23a,
       author = {{McKinney}, Jed and {Manning}, Sinclaire M. and {Cooper}, Olivia R. and {Long}, Arianna S. and {Akins}, Hollis and {Casey}, Caitlin M. and {Faisst}, Andreas L. and {Franco}, Maximilien and {Hayward}, Christopher C. and {Lambrides}, Erini and {Magdis}, Georgios and {Whitaker}, Katherine E. and {Yun}, Min and {Champagne}, Jaclyn B. and {Drakos}, Nicole E. and {Gentile}, Fabrizio and {Gillman}, Steven and {Gozaliasl}, Ghassem and {Ilbert}, Olivier and {Jin}, Shuowen and {Koekemoer}, Anton M. and {Kokorev}, Vasily and {Liu}, Daizhong and {Rich}, R. Michael and {Robertson}, Brant E. and {Valentino}, Francesco and {Weaver}, John R. and {Zavala}, Jorge A. and {Allen}, Natalie and {Kartaltepe}, Jeyhan S. and {McCracken}, Henry Joy and {Paquereau}, Louise and {Rhodes}, Jason and {Shuntov}, Marko and {Toft}, Sune},
        title = "{A Near-infrared-faint, Far-infrared-luminous Dusty Galaxy at z {\ensuremath{\sim}} 5 in COSMOS-Web}",
      journal = {\apj},
         year = 2023,
        month = oct,
       volume = {956},
       number = {2},
          eid = {72},
        pages = {72},
          doi = {10.3847/1538-4357/acf614},
archivePrefix = {arXiv},
       eprint = {2304.07316},
 primaryClass = {astro-ph.GA},
       adsurl = {https://ui.adsabs.harvard.edu/abs/2023ApJ...956...72M}
}

@ARTICLE{Gardner2023,
       author = {{Gardner}, Jonathan P. and {Mather}, John C. and {Abbott}, Randy and {Abell}, James S. and {Abernathy}, Mark and {Abney}, Faith E. and {Abraham}, John G. and {Abraham}, Roberto and {Abul-Huda}, Yasin M. and {Acton}, Scott and {Adams}, Cynthia K. and {Adams}, Evan and {Adler}, David S. and {Adriaensen}, Maarten and {Aguilar}, Jonathan Albert and {Ahmed}, Mansoor and {Ahmed}, Nasif S. and {Ahmed}, Tanjira and {Albat}, R{\"u}deger and {Albert}, Lo{\"\i}c and {Alberts}, Stacey and {Aldridge}, David and {Allen}, Mary Marsha and {Allen}, Shaune S. and {Altenburg}, Martin and {Altunc}, Serhat and {Alvarez}, Jose Lorenzo and {{\'A}lvarez-M{\'a}rquez}, Javier and {Alves de Oliveira}, Catarina and {Ambrose}, Leslie L. and {Anandakrishnan}, Satya M. and {Andersen}, Gregory C. and {Anderson}, Harry James and {Anderson}, Jay and {Anderson}, Kristen and {Anderson}, Sara M. and {Aprea}, Julio and {Archer}, Benita J. and {Arenberg}, Jonathan W. and {Argyriou}, Ioannis and {Arribas}, Santiago and {Artigau}, {\'E}tienne and {Arvai}, Amanda Rose and {Atcheson}, Paul and {Atkinson}, Charles B. and {Averbukh}, Jesse and {Aymergen}, Cagatay and {Bacinski}, John J. and {Baggett}, Wayne E. and {Bagnasco}, Giorgio and {Baker}, Lynn L. and {Balzano}, Vicki Ann and {Banks}, Kimberly A. and {Baran}, David A. and {Barker}, Elizabeth A. and {Barrett}, Larry K. and {Barringer}, Bruce O. and {Barto}, Allison and {Bast}, William and {Baudoz}, Pierre and {Baum}, Stefi and {Beatty}, Thomas G. and {Beaulieu}, Mathilde and {Bechtold}, Kathryn and {Beck}, Tracy and {Beddard}, Megan M. and {Beichman}, Charles and {Bellagama}, Larry and {Bely}, Pierre and {Berger}, Timothy W. and {Bergeron}, Louis E. and {Bernier}, Antoine-Darveau and {Bertch}, Maria D. and {Beskow}, Charlotte and {Betz}, Laura E. and {Biagetti}, Carl P. and {Birkmann}, Stephan and {Bjorklund}, Kurt F. and {Blackwood}, James D. and {Blazek}, Ronald Paul and {Blossfeld}, Stephen and {Bluth}, Marcel and {Boccaletti}, Anthony and {Boegner}, Jr., Martin E. and {Bohlin}, Ralph C. and {Boia}, John Joseph and {B{\"o}ker}, Torsten and {Bonaventura}, N. and {Bond}, Nicholas A. and {Bosley}, Kari Ann and {Boucarut}, Rene A. and {Bouchet}, Patrice and {Bouwman}, Jeroen and {Bower}, Gary and {Bowers}, Ariel S. and {Bowers}, Charles W. and {Boyce}, Leslye A. and {Boyer}, Christine T. and {Boyer}, Martha L. and {Boyer}, Michael and {Boyer}, Robert and {Bradley}, Larry D. and {Brady}, Gregory R. and {Brandl}, Bernhard R. and {Brannen}, Judith L. and {Breda}, David and {Bremmer}, Harold G. and {Brennan}, David and {Bresnahan}, Pamela A. and {Bright}, Stacey N. and {Broiles}, Brian J. and {Bromenschenkel}, Asa and {Brooks}, Brian H. and {Brooks}, Keira J. and {Brown}, Bob and {Brown}, Bruce and {Brown}, Thomas M. and {Bruce}, Barry W. and {Bryson}, Jonathan G. and {Bujanda}, Edwin D. and {Bullock}, Blake M. and {Bunker}, A.~J. and {Bureo}, Rafael and {Burt}, Irving J. and {Bush}, James Aaron and {Bushouse}, Howard A. and {Bussman}, Marie C. and {Cabaud}, Olivier and {Cale}, Steven and {Calhoon}, Charles D. and {Calvani}, Humberto and {Canipe}, Alicia M. and {Caputo}, Francis M. and {Cara}, Mihai and {Carey}, Larkin and {Case}, Michael Eli and {Cesari}, Thaddeus and {Cetorelli}, Lee D. and {Chance}, Don R. and {Chandler}, Lynn and {Chaney}, Dave and {Chapman}, George N. and {Charlot}, S. and {Chayer}, Pierre and {Cheezum}, Jeffrey I. and {Chen}, Bin and {Chen}, Christine H. and {Cherinka}, Brian and {Chichester}, Sarah C. and {Chilton}, Zachary S. and {Chittiraibalan}, Dharini and {Clampin}, Mark and {Clark}, Charles R. and {Clark}, Kerry W. and {Clark}, Stephanie M. and {Claybrooks}, Edward E. and {Cleveland}, Keith A. and {Cohen}, Andrew L. and {Cohen}, Lester M. and {Col{\'o}n}, Knicole D. and {Coleman}, Benee L. and {Colina}, Luis and {Comber}, Brian J. and {Comeau}, Thomas M. and {Comer}, Thomas and {Conde Reis}, Alain and {Connolly}, Dennis C. and {Conroy}, Kyle E. and {Contos}, Adam R. and {Contreras}, James and {Cook}, Neil J. and {Cooper}, James L. and {Cooper}, Rachel Aviva and {Correia}, Michael F. and {Correnti}, Matteo and {Cossou}, Christophe and {Costanza}, Brian F. and {Coulais}, Alain and {Cox}, Colin R. and {Coyle}, Ray T. and {Cracraft}, Misty M. and {Crew}, Keith A. and {Curtis}, Gary J. and {Cusveller}, Bianca and {Da Costa Maciel}, Cleyciane and {Dailey}, Christopher T. and {Daugeron}, Fr{\'e}d{\'e}ric and {Davidson}, Greg S. and {Davies}, James E. and {Davis}, Katherine Anne and {Davis}, Michael S. and {Day}, Ratna and {de Chambure}, Daniel and {de Jong}, Pauline and {De Marchi}, Guido and {Dean}, Bruce H. and {Decker}, John E. and {Delisa}, Amy S. and {Dell}, Lawrence C. and {Dellagatta}, Gail},
        title = "{The James Webb Space Telescope Mission}",
      journal = {\pasp},
         year = 2023,
        month = jun,
       volume = {135},
       number = {1048},
          eid = {068001},
        pages = {068001},
          doi = {10.1088/1538-3873/acd1b5},
archivePrefix = {arXiv},
       eprint = {2304.04869},
 primaryClass = {astro-ph.IM},
       adsurl = {https://ui.adsabs.harvard.edu/abs/2023PASP..135f8001G}
}

@ARTICLE{2023ApJ...958...36S,
       author = {{Smail}, Ian and {Dudzeviciute}, Ugne and {Gurwell}, Mark and others},
        title = "{Hidden Giants in JWST's PEARLS: An Ultramassive z = 4.26 Submillimeter Galaxy that Is Invisible to HST}",
      journal = {\apj},
         year = 2023,
        month = nov,
       volume = {958},
       number = {1},
          eid = {36},
        pages = {36},
          doi = {10.3847/1538-4357/acf931},
       adsurl = {https://ui.adsabs.harvard.edu/abs/2023ApJ...958...36S}
}

@ARTICLE{2025ApJ...982..125C,
       author = {{Cooper}, Olivia R. and {Brammer}, Gabriel and {Heintz}, K.~E. and others},
        title = "{RUBIES: JWST/NIRSpec Resolves Evolutionary Phases of Dusty Star-forming Galaxies at $z\sim2$}",
      journal = {\apj},
         year = 2025,
        month = apr,
       volume = {982},
       number = {2},
          eid = {125},
        pages = {125},
          doi = {10.3847/1538-4357/adb8e1},
       adsurl = {https://ui.adsabs.harvard.edu/abs/2025ApJ...982..125C}
}

@ARTICLE{2026A&A...705A.214L,
       author = {{Lagache}, G. and {Xiao}, M. and {Beelen}, A. and others},
        title = "{Overdense Fireworks in GOODS-N: Unveiling a Record Number of Massive Dusty Star-forming Galaxies at $z\sim5.2$ with the N2CLS}",
      journal = {\aap},
         year = 2026,
        month = jan,
       volume = {705},
          eid = {A214},
        pages = {A214},
          doi = {10.1051/0004-6361/202555947},
       adsurl = {https://ui.adsabs.harvard.edu/abs/2026A&A...705A.214L}
}

@ARTICLE{2025ApJ...980...12S,
       author = {{Sun}, Fengwu and {Wang}, Feige and {Yang}, Jinyi and others},
        title = "{A SPectroscopic Survey of Biased Halos in the Reionization Era (ASPIRE): Spectroscopically Complete Census of Obscured Cosmic Star Formation Rate Density at z = 4--6}",
      journal = {\apj},
         year = 2025,
        month = feb,
       volume = {980},
       number = {1},
          eid = {12},
        pages = {12},
          doi = {10.3847/1538-4357/ad9d0e},
archivePrefix = {arXiv},
       eprint = {2412.06894},
 primaryClass = {astro-ph.GA},
       adsurl = {https://ui.adsabs.harvard.edu/abs/2025ApJ...980...12S}
}

@ARTICLE{2024ApJ...973...25K,
       author = {{Kamieneski}, Patrick S. and {Frye}, Brenda L. and {Windhorst}, Rogier A. and others},
        title = "{Birds of a Feather: Resolving Stellar Mass Assembly with JWST/NIRCam in a Pair of Kindred $z\sim2$ Dusty Star-forming Galaxies Lensed by the PLCK G165.7+67.0 Cluster}",
      journal = {\apj},
         year = 2024,
        month = sep,
       volume = {973},
       number = {1},
          eid = {25},
        pages = {25},
          doi = {10.3847/1538-4357/ad5d59},
       adsurl = {https://ui.adsabs.harvard.edu/abs/2024ApJ...973...25K}
}

\medskip
{\footnotesize
\noindent\textbf{Affiliations}\\
\noindent $^1$Department of Astronomy, School of Science, Westlake University, Hangzhou, Zhejiang 310030, P.\ R.\ China\\
$^2$SKA Observatory, Jodrell Bank, SK11 9FT, UK\\
$^3$Institute of Astronomy and Astrophysics, Academia Sinica, 11F of Astronomy-Mathematics Building, No. 1, Section 4, Roosevelt Road, Taipei 106319, Taiwan, R.O.C.\\
$^4$Instituto de Estudios Astrof\'{\i}sicos, Facultad de Ingenier\'{\i}a y Ciencias, Universidad Diego Portales, Avenida Ej\'ercito Libertador 441, Santiago, Chile\\
$^5$Department of Physics and Astronomy, Chalmers University of Technology, SE-412 96 Gothenburg, Sweden\\
$^6$Kapteyn Astronomical Institute, University of Groningen, P.O. Box 800, 9700 AV Groningen, The Netherlands\\
$^7$Sorbonne Universit\'e, UPMC Paris 6 and CNRS, UMR 7095, Institut d'Astrophysique de Paris, 98b bd. Arago, 75014 Paris, France\\
$^8$International Centre for Radio Astronomy Research, University of Western Australia, ICRAR M468, 35 Stirling Highway, Crawley, WA 6009, Australia\\
$^9$IPAC, California Institute of Technology, 1200 E. California Blvd., Pasadena, CA 91125, USA\\
$^{10}$Center for Frontier Science, Chiba University, 1-33 Yayoi-cho, Inage-ku, Chiba 263-8522, Japan\\
$^{11}$INAF, Osservatorio di Astrofisica e Scienza dello Spazio, Via P. Gobetti 93/3, 40129 Bologna, Italy\\
$^{12}$Centro de Astrobiolog\'ia (CAB), CSIC-INTA, Carretera de Ajalvir km 4, Torrej\'on de Ardoz, E-28850 Madrid, Spain\\
$^{13}$Leiden Observatory, Leiden University, P.O. Box 9513, 2300 RA Leiden, The Netherlands\\
$^{14}$Instituto de Astronom\'ia y Ciencias Planetarias, Universidad de Atacama, Copayapu 485, Copiap\'o 1531772, Chile\\
$^{15}$SISSA, Via Bonomea 265, Trieste, Italy\\
$^{16}$Italian National Institute of Astrophysics--Astronomical Observatory of Trieste, Via G. Tiepolo 11, 34143 Trieste, Italy\\
$^{17}$Department of Astronomy, University of Michigan, 1085 South University Avenue, Ann Arbor, MI 48109, USA\\
$^{18}$Pole Scientific, University Catholic of Lyon--University of Lyon, 10 Place des Archives, 69002 Lyon, France\\
$^{19}$Institut f\"ur Astrophysik, Universit\"at zu K\"oln, Z\"ulpicher Stra{\ss}e 77, D-50937 K\"oln, Germany\\
$^{20}$Cluster of Excellence ``Our Dynamic Universe'' (DYNAVERSE), Germany\\
$^{21}$Department of Astrophysics, University of Vienna, T\"urkenschanzstrasse 17, 1180 Vienna, Austria
}
\end{document}